\documentclass[aip,apl,reprint,superscriptaddress,amsmath,amssymb]{revtex4-1}

\usepackage[]{graphicx}
\usepackage{lmodern}
\usepackage{subfigure}
\usepackage[version=3]{mhchem} 

\usepackage[utf8]{inputenc}
\usepackage{lmodern}
\usepackage[T1]{fontenc}
\usepackage{amsmath, amssymb}
\usepackage{hyperref}
\usepackage{xcolor}
\usepackage{notes2bib}

\newcommand{\beq}{\begin{equation}\begin{aligned}}
\newcommand{\eeq}{\end{aligned}\end{equation}}

\definecolor{linkcol}{rgb}{0,0,0.4} 
\definecolor{citecol}{rgb}{0.5,0,0} 

\hypersetup{
bookmarksopen=true,
pdftitle="Hole Transport in Exfoliated Monolayer MoS2",
pdfauthor="Evgeniy Ponomarev et al",
pdfsubject="arxiv", 
pdfstartview={FitH},    
pdfmenubar=true, 
pdfhighlight=/O, 
colorlinks=true, 
pdfpagemode=UseNone, 
pdfpagelayout=SinglePage, 
pdffitwindow=true, 
linkcolor=linkcol, 
citecolor=linkcol, 
urlcolor=blue
}

\begin{document}

\title{\texorpdfstring{\Large Hole Transport in Exfoliated Monolayer MoS$_{2}$ \\} {Hole Transport in Exfoliated Monolayer MoS$_{2}$}}

	\author{Evgeniy Ponomarev}
\affiliation{DQMP, Université de Genève, 24 quai Ernest Ansermet, CH-1211, Geneva, Switzerland}
\affiliation{GAP, Université de Genève, 24 quai Ernest Ansermet, CH-1211, Geneva, Switzerland}

\author{Árpád Pásztor}
\affiliation{DQMP, Université de Genève, 24 quai Ernest Ansermet, CH-1211, Geneva, Switzerland}

\author{Adrien Waelchli}
\affiliation{DQMP, Université de Genève, 24 quai Ernest Ansermet, CH-1211, Geneva, Switzerland}

\author{Alessandro Scarfato}
\affiliation{DQMP, Université de Genève, 24 quai Ernest Ansermet, CH-1211, Geneva, Switzerland}

\author{Nicolas Ubrig}
\affiliation{DQMP, Université de Genève, 24 quai Ernest Ansermet, CH-1211, Geneva, Switzerland}
\affiliation{GAP, Université de Genève, 24 quai Ernest Ansermet, CH-1211, Geneva, Switzerland}

\author{Christoph Renner}
\affiliation{DQMP, Université de Genève, 24 quai Ernest Ansermet, CH-1211, Geneva, Switzerland}

\author{Alberto F. Morpurgo}
\email{Alberto.Morpurgo@unige.ch}
\affiliation{DQMP, Université de Genève, 24 quai Ernest Ansermet, CH-1211, Geneva, Switzerland}
\affiliation{GAP, Université de Genève, 24 quai Ernest Ansermet, CH-1211, Geneva, Switzerland}



\date{\today}

\begin{abstract}
	Ideal monolayers of common semiconducting transition metal dichalcogenides (TMDCs) such as MoS$_2$, WS$_2$, MoSe$_2$, and WSe$_2$ possess many similar electronic properties. As it is the case for all semiconductors, however, the physical response of these systems is strongly determined by defects in a way specific to each individual compound. Here we investigate the ability of exfoliated monolayers of these TMDCs to support high-quality, well-balanced ambipolar conduction, which has been demonstrated for WS$_2$, MoSe$_2$, and WSe$_2$, but not for MoS$_2$. Using ionic-liquid gated transistors we show that, contrary to WS$_2$, MoSe$_2$, and WSe$_2$, hole transport in exfoliated MoS$_2$ monolayers is systematically anomalous, exhibiting a maximum in conductivity at negative gate voltage (V$_G$) followed by a suppression of up to 100 times upon further increasing V$_G$. To understand the origin of this difference we have performed a series of experiments including the comparison of hole transport in MoS$_2$ monolayers and thicker multilayers, in exfoliated and CVD-grown monolayers, as well as gate-dependent optical measurements (Raman and photoluminescence) and scanning tunneling imaging and spectroscopy. In agreement with existing {\it ab-initio} calculations, the results of all these experiments are  consistently explained in terms of defects associated to chalcogen vacancies that only in MoS$_2$ monolayers -- but not in thicker MoS$_2$ multilayers nor in monolayers of the other common semiconducting TMDCs -- create in-gap states near the top of the valence band that act as strong hole traps. Our results demonstrate the importance of studying systematically how defects determine the properties of 2D semiconducting materials and of developing methods to control them.
\end{abstract}


\maketitle

Extensive studies of monolayers (MLs) of group VI semiconducting transition metal dichalcogenides (TMDCs) have demonstrated that all these materials share many electronic properties. For example, they all have a direct band gap at the K and K` points,\cite{Splendiani2010,Mak2010,Zhao2013} a finite Berry curvature in the K and K` valley \cite{Zeng2012,Mak2012,Cao2012,Jones2013,Mak2014} responsible for the occurrence of the valley Hall effect\cite{Mak2014,Ubrig2017},  an extremely strong spin-orbit coupling (as large as a few hundreds meV in the valence band),\cite{Zhu2011,Xiao2012,Komider2013} very large exciton binding energies due to the reduced screening characteristic of 2D systems,\cite{Peimyoo2013,He2014,Chernikov2014} stable trion excitations\cite{Mak2013,Shang2015}, and more. Differences are also present, such as the relative sign of the spin orientation at the conduction band minimum (CBM) and valence band maximum (VBM) -- the same for Mo-based compounds and opposite for W-based ones -- that leads to a different temperature dependence of the measured photoluminescence.\cite{Ye2014,Zhang2015} Although important for specific physical phenomena, these differences  mostly concern more subtle aspects of the electronic properties.

Having access to a broad class of semiconducting 2D materials with many similar properties is very attractive because -- for instance -- it facilitates the realization of van der Waals heterostructures obtained by stacking two or more monolayers of different TMDCs on top of each other. It should be realized, however, that these considerations do not take into account that real materials unavoidably contain defects that are specific to each individual compound, and that drastically affect their electronic response. This is certainly the case for the systems considered here, since in semiconductors defects generally determine crucial characteristics such as the work function,\citep{Addou2015} the position of electrochemical potential,\cite{McDonnell2014,Mahjouri-Samani2016} the transport properties (\textit{e.g.}, the carrier mobility \cite{Qiu2013,Hong2015}), the rate of non-radiative electron-hole recombination, \cite{Wang2015} \textit{etc}. That is why an increasing research effort is currently being devoted to the investigation of defects present in all types of semiconducting 2D materials, whose identification, understanding and control will be necessary if these systems will eventually be employed in technological applications (not to mention the possibility to exploit  new functionalities that are sometimes offered by defects in 2D materials, such as -- for instance -- their ability to act as single photon emitters\cite{Koperski2015,Srivastava2015,He2015,Chakraborty2015,Bourrellier2016,Grosso2017}).

One important aspect that is seemingly common to semiconducting TMDC monolayers is their ability to support well-balanced ambipolar transport. Measurements done on suitable field-effect transistor (FET) devices upon sweeping the gate voltage show that an equally good conductivity is found in a same monolayer irrespective of whether the chemical potential is in the conduction or in the valence band. This feature is particularly relevant for the realization of opto-electronic devices, since it is the ability to transport simultaneously electrons and holes that allows the controlled generation of electroluminescence from electron-hole recombination or the conversion of light into an electrical signal. Surprisingly, however, if we look at experiments reported on exfoliated monolayers of common group VI semiconducting TMDCs, well-balanced ambipolar transport has been observed in MoSe$_{2}$,\cite{Onga2016} WSe$_{2}$,\cite{Allain2014} and WS$_{2}$,\cite{Jo2014} but not in MoS$_{2}$. This is unexpected both because exfoliated MoS$_{2}$ monolayers are probably the most studied among these compounds (which is why they are often used to benchmark other 2D semiconductors) and because for thick exfoliated MoS$_{2}$ multilayers excellent ambipolar conduction is routinely observed. \cite{Zhang2012,Zhang2013}

Motivated by these considerations, here we investigate ambipolar conduction in exfoliated MoS$_{2}$ monolayers using ionic liquid-gated FETs and demonstrate a systematic and reproducible anomalous behavior of transport upon hole accumulation. Specifically, after the onset of hole conduction a virtually complete suppression of source-drain current is observed in all devices as the gate voltage is biased to shift the chemical potential deeper into the valence band. In contrast, experiments on identical devices realized on exfoliated bi, tri, and tetralayer MoS$_{2}$ show excellent ambipolar conduction, demonstrating that the anomalous hole transport is an inherent characteristic of monolayer MoS$_{2}$ devices. By combining transistor measurements, gate-dependent optical studies (photo-luminescence and Raman spectroscopy), scanning tunneling imaging and spectroscopy, and a thorough analysis of existing studies based on \textit{ab-initio} calculations, we establish the origin of this phenomenon as due to the presence of atomic-scale defects that induce states inside the band-gap of MoS$_{2}$ monolayers, approximately 300-400 meV  above the top of the valence band. These defects -- whose manifestations are consistent with what is expected from sulfur vacancies \cite{Yuan2014,Vancso2016} -- act as traps for holes and prevent hole conduction in exfoliated MoS$_2$ monolayers. The phenomenon is specific to MoS$_2$ monolayers -- and not to monolayers of other semiconducting TMDCs or to MoS$_2$ bilayers/thicker multilayers -- because only in MoS$_2$ monolayers the states created by chalcogen vacancies near the top of the valence band appear to be inside the band gap.

\section*{Results ans discussion}
The most effective way to investigate ambipolar transport in TMDC monolayers is by integrating them into a FET employing an ionic liquid top gate, as illustrated schematically in Figure 1a. The very large capacitance of the ionic gate allows the electrochemical potential to be shifted over a large range -- from deep in the  conduction band to deep in the valence band -- as demonstrated experimentally by the ambipolar conduction observed in WS$_{2}$\cite{Jo2014}, WSe$_{2}$ \cite{Allain2014} and MoSe$_{2}$ \cite{Onga2016,Chen2017} monolayers. For FETs using exfoliated monolayers of MoS$_{2}$, however, a similar observation has not been reported, and our measurements show that an unexpected behavior indeed occurs when the gate is biased to shift the electrochemical potential into the valence band. Specifically, Figure 1b shows transfer curves (source-drain current I$_{SD}$ as a function of gate voltage V$_{G}$ at a finite applied source-drain bias $V_{SD}$) measured on two of our devices representative of the behavior of eight nominally identical devices that we investigated. Upon increasing V$_{G}$ above the electron threshold voltage ($V_{th}^{e}$) carriers are accumulated in the conduction band and the source-drain current increases steeply (the estimated value of electron density  $n _{e}$  at the largest V$_G$ value reached approximately  $n _{e} = C_{*}(V_{G}-V_{th}^{e})/e=7\cdot10^{12}$ cm$^{-2}$ where $C_{*}= 7 \mu Fcm^{-2}$ is the capacitance per unit of the ionic liquid \cite{Braga2012,Jo2014}; values as large as $n _{e} = 2-3 \cdot10^{14}$ cm$^{-2}$ can be obtained at larger positive V$_G$ values \cite{Costanzo2016}). This is the behavior that is commonly expected in a field-effect transistor.

\begin{figure}
	\centering
	\includegraphics[width =.5\textwidth]{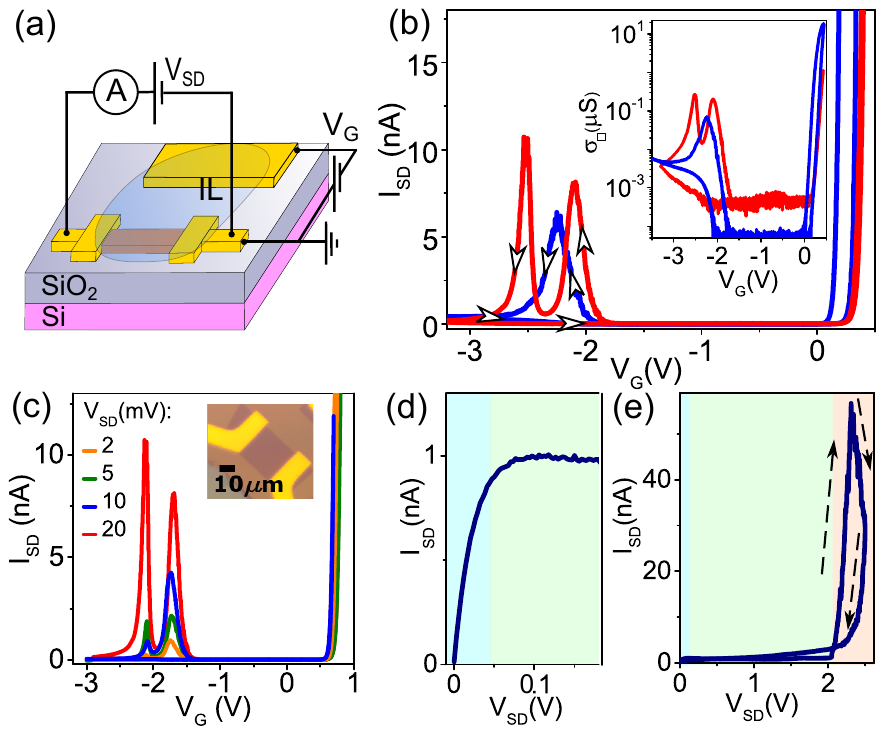}
	\caption{Anomalous hole transport in FETs based on exfoliated monolayer MoS$_{2}$. (a) Schematics of an ionic-liquid gated FET based on monolayer MoS$_2$. (b) Representative FET transfer curves (\textit{i.e.}, I$_{SD}$ versus V$_{G}$ at fixed V$_{SD}$) showing electron transport for V$_{G}$ > 0.3V and an anomalous transport behavior upon hole accumulation: for V$_{G}$ < -1.9V, after an initial increase, the source-drain current decreases to vanishingly small values if V$_{G}$ is swept to larger negative values. Depending on the device either one (blue curve, V$_{SD}$ = 100mV) or two (red curve, V$_{SD}$ = 20mV) peaks in I$_{SD}$ are seen upon hole accumulation. In all cases, as V$_{G}$ is swept back from large negative values, I$_{SD}$ remains low (the arrows indicate how V$_{G}$ is swept). (Inset) Same data shown in panel (b) plotted in terms of conductivity  $\sigma_{\square}$. (c) I$_{SD}$-vs-V$_{G}$ measured for different values of V$_{SD}$: the measurements, done in sequence, show that the current suppression at large negative V$_{G}$ does not originate from a permanent device degradation (the inset shows an optical microscope image of a device). In (b) and (c) the relative V$_G$-shift visible in the transfer curves originates from bias stress caused by prolonged measurements done on this specific device. (d,e) Source-drain current I$_{SD}$ as a function of source-drain bias V$_{SD}$. Panel (d) focuses on  the linear and saturation regimes (respectively shaded in light blue and light green). Panel (e) shows a much larger V$_{SD}$ range; the large increase in current for V$_{SD}$ > 2V is due to the ambipolar injection regime (light orange shaded region) in which electrons and holes are injected at opposite contacts. The abrupt decrease in I$_{SD}$, occurring just after the onset of hole accumulation, is a manifestation of the same effect seen in panel (b).}
	
\end{figure}

As V$_{G}$ is swept past the onset of hole transport, however, we observe that after an initial increase, I$_{SD}$ exhibits a maximum followed by a steep drop to a virtually vanishing current, with the conductivity $\sigma_{\square}$ dropping by approximately two orders of magnitude (see the inset in Figure 1b). Depending on the specific device either one or two peaks in I$_{SD}$ are observed, as shown by the red and blue curves in Figure 1b. With any further increase of gate voltage to more negative values $I_{SD}$ stays low, and remains low as the gate voltage is swept back to $V_G=$ 0V. The phenomenon is reversible upon re-cycling $V_G$ from zero to large negative values and back, implying that the observed behavior is not due to a permanent degradation of the device. This can be concluded from Figure 1c, which shows transfer curves acquired successively one after the other, by sweeping $V_G$, for increasingly large V$_{SD}$ values. The phenomenon was observed in FET devices with source and drain contacts made of two different materials (Au and Pt), indicating that the current suppression -- and the corresponding anomalous behavior of transport upon hole accumulation -- is not due to a contact effect (FET transfer curves measured on devices with Pt contacts are shown in figure S1).

The anomalous behavior of hole conduction in MoS$_{2}$ monolayers is also clearly visible in the FET output characteristics, \textit{i.e.}, when measuring the source-drain current I$_{SD}$ as a function of source-drain bias V$_{SD}$ at a fixed gate voltage $V_G$. At low positive  V$_{SD}$ values, with the gate biased to have electron accumulation, the usual transistor behavior is observed, with  $I_{SD}$ increasing linearly until the onset of the saturation regime, occurring at V$_{SD} \simeq$ V$_{G}$-V$_{th}^{e}$, past which $I_{SD}$ stays constant (Figure 1d). As $V_{SD}$ is increased up to much higher values, $I_{SD}$ starts to grow again rapidly (Figure 1e), because the applied source-drain bias causes the channel potential to reverse its polarity and to exceed the threshold for hole accumulation near the drain contact. This is the so-called ambipolar injection regime in which the steep increase in I$_{SD}$ is due to holes injected from the drain contact. In the experiment we see that, sweeping $V_{SD}$ past the point when $I_{SD}$ starts to increase (orange shaded area in Figure 1e) causes a rapid drop of current, and the current stays low as $V_{SD}$ is swept back. We conclude that, irrespectively of the way in which holes are accumulated in the MoS$_{2}$ monolayer channel -- either by sweeping the gate voltage or the source-drain bias -- an unexpected anomalous suppression of I$_{SD}$ is observed in all cases.

The observed phenomenon is inherent to monolayers, and is absent in ionic liquid gated FETs realized with MoS$_2$ bilayers or thicker multilayers. Figure 2 shows the transfer curves of devices fabricated on bi, tri, and tetra layer MoS$_{2}$, in which the measured $I_{SD}$ is excellently balanced (\textit{i.e.}, the magnitude of the measured current is approximately the same upon electron and hole accumulation), with no indication of any anomalous behavior. These observations explain why earlier experiments on ionic liquid gated FETs realized using thick (essentially bulk-like) exfoliated MoS$_{2}$ crystals\cite{Zhang2012,Zhang2013} have reported high-quality ambipolar transport, despite the absence of hole current in monolayers.

\begin{figure}
	\centering
	\includegraphics[width =0.5\textwidth]{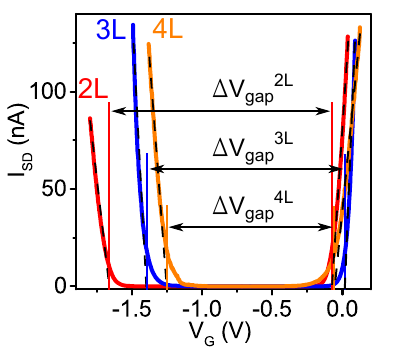}
	\caption{Ambipolar transport in MoS$_{2}$ multilayers. Source-drain current I$_{SD}$ as a function of gate voltage V$_{G}$ for FETs fabricated on MoS$_{2}$ multilayers, containing respectively two, three, and four monolayers. Well balanced ambipolar transport is seen in all cases. The black dashed lines are the extrapolation of I$_{SD}$ to 0 nA, as required to extract the values of threshold voltage (the thresholds for hole and electron conductions in the layers of different thicknesses are indicated by the vertical solid lines of the corresponding colors). The band gap is obtained from  E$_{gap }$ = e (V$ ^{e}_{th}$ - V$ ^{h}_{th}$).}
	
\end{figure}

Note that, as the thickness of the MoS$_2$ layer is increased, the range of gate voltages in which the current vanishes (\textit{i.e.}, in which the electrochemical potential is located inside the band gap) decreases. This is because ionic liquid gating used on systems with a small density of states, such as in the gap of a semiconductor, has spectroscopic capabilities\cite{Braga2012,Jo2014,Lezama2014} and the difference in $V_G$ between the threshold voltage for electron and hole conduction is a direct measure of the band gap (to increase the precision with which the gap is determined, data should be plotted as a function of a reference potential -- which avoids effects caused by a possible voltage drop at the gate-liquid interface -- but in high quality devices the reference potential and the gate voltage nearly coincide). We conclude from these measurements that the band gap is $\Delta$V$_{gap }^{2L}$ = 1.6 eV for MoS$_2$ bilayers,  $\Delta$V$_{gap }^{3L}$ = 1.4 eV for trilayers, and $\Delta$V$_{gap }^{4L}$ = 1.25 eV for tetralayers, in all cases with a precision of approximately 10\%. Importantly, since the threshold voltage determines the energy of accumulation of individual charge carriers in the respective band, the values extracted are the actual band gaps of the different multilayers -- as measured for instance from scanning tunneling spectroscopy \cite{Huang2015a}-- and not the exciton energies which are commonly measured in optical experiments. Irrespective of these considerations, we re-iterate that these experiments unambiguously show that the suppression of conductivity under hole accumulation is a distinct property of exfoliated MoS$_{2}$ monolayers.

The systematic anomalous hole transport behavior observed in MoS$_2$ monolayer devices is unexpected and an explanation is called for. In view of the very large negative gate voltage required to populate the valence band, one may wonder whether the correspondingly large electric field (estimated by dividing the maximum applied gate voltage by the thickness of the ionic liquid double layer, and reaching up to several tens of MV/cm) can actually affect the crystalline structure of the material. Indeed, such a drastic structural effect has been reported previously for MoTe$_{2}$ \cite{Li2016,Wang2017}, and its occurrence in monolayer MoS$_2$ cannot be excluded a priori. To address this question we performed Raman and photoluminescence spectroscopy of ionic-liquid gated MoS$_{2}$ monolayers \textit{in-situ} , in the presence of a large negative gate voltage (see schematics in Figure 3a). Selected Raman spectra collected at different $V_G$ values, from 0 V up to -3.3 V (see Figure 3b), show no change in the characteristic E$_{2g}$ and A$_{1g}$ modes of 2H MoS$_{2}$ throughout the $V_G$ range investigated (the same is true for other parts of the spectrum that we looked at). Results from photoluminescence (PL) spectroscopy are shown in Figure 3c, with a clear peak centered at about 670nm that originates from the A-exciton recombination in MoS$_{2}$ and represents a characteristic signature of MoS$_2$ monolayers. Remarkably, the peak (broadened by large potential fluctuations generated by the ionic liquid gate) remains unchanged, irrespective of the applied gate voltage: the complete photoluminescence map together with the corresponding FET transfer curve are represented in figure S2 (in these measurements the V$_G$-range of the FET transfer curve is shifted relative to the one shown in Figure 1b due to bias stress and the anomalous hole transport is observed at somewhat more negative values of $V_G$). The observed insensitivity of Raman and PL measurements to the gate voltage indicates the absence of significant structural changes in the MoS$_2$ monolayers, and allows us to exclude that the reentrance of the insulating state observed at large negative $V_G$ has a structural origin.

\begin{figure}
	\centering
	\includegraphics[width =0.5\textwidth]{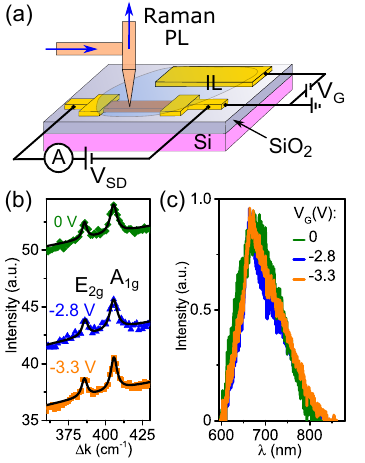}
	\caption{\textit{In-situ} optical characterization of an ionic liquid-gated  MoS$_{2}$ monolayer FET. (a) Measurement scheme to perform gate dependent photoluminescence and Raman spectroscopy on a MoS$_{2}$ monolayer. (b) Raman spectra showing the characteristic vibrational modes of MoS$_{2}$ monolayers (E$_{2g}$ and A$_{1g}$) for different V$_{G}$ values. No significant changes in position or line-width of these modes is seen upon applying a large negative gate voltage, corresponding to values for which holes are accumulated in the channel. (c) Photoluminescence spectra for different values of V$_{G}$ showing that the A-exciton peak of MoS$_{2}$ monolayer remains unchanged.}
	\label {fig:ref 03}
\end{figure}

What makes the suppression of hole current in exfoliated MoS$_{2}$ monolayers even more surprising is that FETs realized on large-area monolayers grown by chemical vapor deposition (CVD) do exhibit good ambipolar characteristics.\cite{Ponomarev2015}  The top panel of figure 4 shows the conductivity ($\sigma_{\square}$) of a device realized on such a CVD monolayer (blue curve) that, in contrast to what is measured on a device realized using an exfoliated monolayer (green curve), exhibits rather conventional ambipolar conduction. From the data, we first extract the threshold voltage for electrons (which coincides for CVD-grown and exfoliated monolayers) and holes (which can only be determined for the device realized on the CVD-grown monolayer); then we compare transport in the electron and hole subthreshold regimes by looking at the source-drain current $I_{SD}$ on a logarithmic scale (Figure 4, bottom panel). In the electron sub-threshold regime (light-blue shaded region in Figure 4) the conductivity is higher and the subthreshold slope is less steep for the device realized on the CVD-grown monolayer. This behavior is due to a broader tail of disorder-induced localized states that in CVD-grown monolayers is present inside the band gap near the bottom of the conduction band.  It is expected, because CVD-grown monolayers are generally more disordered than exfoliated ones, due to a larger density of grain boundaries.

\begin{figure}
	\centering
	\includegraphics[width =0.45\textwidth]{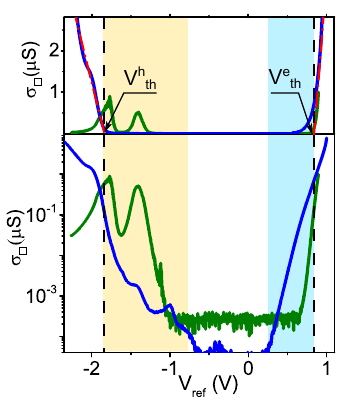}
	\caption{Comparison of ambipolar transport in FETs based on CVD-grown and exfoliated MoS$_{2}$ monolayers. Top panel: Conductivity $\sigma_{\square}$ as a function of reference voltage V$_{ref}$ for CVD-grown (blue curve) and exfoliated (green curve)  MoS$_{2}$ monolayer devices. The red dashed-dotted lines are an extrapolation of the conductivity  $\sigma_{\square}$ measured in the linear regime to 0 S, from which we determine the threshold voltages for electrons and holes. Bottom panel: Semilogarithmic plots of the data shown in the top panel. The subthreshold regimes for electrons and holes correspond to the light blue and the light yellow shaded areas. The higher conductivity of CVD-grown monolayer FET in the electron subthreshold regime originates from the larger density of structural defects present. In the hole subthreshold region the exfoliated monolayer FET exhibits conductivity almost two orders of magnitude higher than the CVD-grown one, as well as a very different qualitative behavior of transport as compared to electrons (subthreshold electron transport is featureless, whereas for holes distinct  peaks in conductivity are seen). These observations are indicative of the very different nature of disorder affecting electron and hole transport in the respective subthreshold regimes.}
	\label {fig:ref 04}
\end{figure}

A very different behavior is observed in the hole subthreshold regime (light yellow shaded region in Figure 4): the current enhancement extends much more deeply into the gap and it is much larger (by nearly two orders of magnitude) for exfoliated monolayers as compared to CVD-grown material.  Furthermore, the very pronounced broad peaks seen in the source-drain current measured on exfoliated monolayers -- also present at comparable energy in CVD-grown material, albeit much less pronouncedly -- seem to indicate the presence of discrete energy states. These observations indicate that the nature of disorder affecting states near the top of the valence band is qualitatively different from that influencing states near the bottom of the conduction band. Specifically, finding that the effect of disorder near the top of the valence band is much stronger in exfoliated MoS$_2$ monolayers  despite their superior structural quality indicates that the origin of the effect is not due to the presence of grain boundaries, but rather to other kind of defects. These observations also strongly suggest that disorder at energies near the top of the valence band has the same origin in exfoliated and CVD-grown monolayers, since the sub-gap states are present in the same range of energies in the two cases.

To investigate the nature of the discrete in-gap energy states near the top of the valence band we employ scanning tunneling microscopy (STM) and scanning tunneling spectroscopy (STS). These techniques provide detailed information on the local crystalline and electronic structure of a material down to the atomic scale. STM measurements were performed on a CVD-grown monolayer deposited on a dielectric substrate (SiO$_{2}$/Si), made conductive by electrostatic accumulation of charge carriers (\textit{i.e.}, by integrating it into a FET device; the set-up scheme and FET transfer characteristics are shown in figure 5a and 5b respectively). Under these conditions, with a large positive gate voltage applied to the Si back gate, we can successfully establish stable tunneling conditions, as it is needed to take hiqh-quality images and perform tunneling spectroscopy. Figure 5c demonstrates that atomic resolution can indeed be achieved through the observation of the triangular atomic lattice of MoS$_2$. Atomic resolution is further highlighted by the intense six-fold symmetric peak structure in the image Fourier-transform. These results illustrate how STM measurements performed under an applied gate voltage allow the local properties of the material to be investigated.

\begin{figure}
	\centering
	\includegraphics[width =0.45\textwidth]{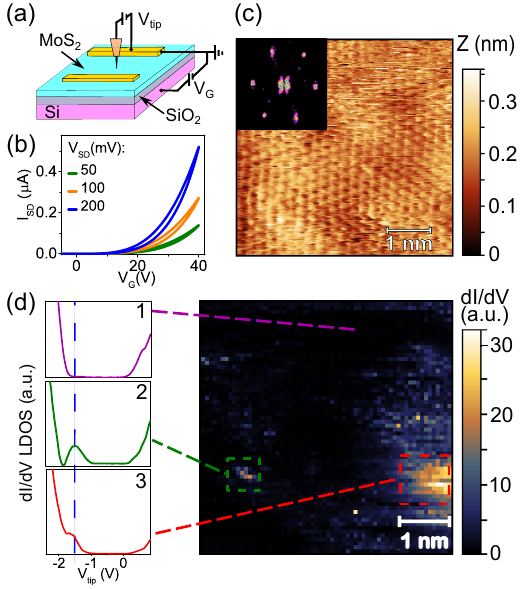}
	\caption{Atomic-scale study of monolayer MoS$_{2}$. (a) Schematic representation of the configuration used to perform STM/STS measurements on a CVD-grown MoS$_{2}$ monolayer. The monolayer is grown on a doped Si substrate acting as a gate covered by a thermally grown SiO$_2$ 285 nm thick layer. STM/STS measurements are done with a positive applied gate voltage that makes the MoS$_2$ monolayer conductive (see the transfer curves shown in (b), measured on the same structure used for STM/STS experiments) (c) Atomic resolution STM image (V$_{tip}$ = 1V, I$_{t}$ = 10pA, V$_{G}$ = 70V) of MoS$_{2}$ monolayer showing the triangular atomic lattice, further highlighted in the inset by the six-fold symmetric structure visible in the 2D Fourier transform of the image. (d) Local STS map showing the differential conductance dI/dV measured at V$_{tip}$ = -1.5V as a function of tip position. Three regions are identified: region 1 (which occupies most of the image) with no states inside the band gap (violet STS spectrum); region 2 (delimited by the green dashed line) in which pronounced states are seen in the spectrum, close to the top of the valence band; region 3 (delimited by the red dashed line) with in-gap states visible in the tunneling spectrum close to the top of the valence band that are broader in energy. Note that the sub-nanometer size of regions 2 and 3 indicates that the in-gap states originate from atomic-scale defects and is compatible with sulfur vacancies. The STM image in (c) and the STS map in (d) were collected from different areas of a same sample.}
	\label {fig:ref 05}
\end{figure}

We probe the spatial distribution of the local density of states (LDOS) by acquiring STS spectra over a dense grid. A map obtained in this way is presented in Figure 5d, where we plot the differential tunneling conductance dI/dV measured at a fixed dc bias (V$_{tip}$ = -1.5 V). Three regions can be identified. The first region occupies the majority of the scanned area and is composed of semiconducting spectra (such as spectrum 1, on the left side of the figure) with a band gap value of 2.1 eV, or just slightly larger (which is why the differential conductance measured at V$_{tip}$ = -1.5 V vanishes). This value of the gap confirms that the material studied is indeed a monolayer. In the two other regions -- delimited by the green (region 2) and red (region 3) dashed lines -- additional states are visible, and manifest themselves as an enhanced tunneling conductance inside the gap of monolayer MoS$_2$. Interestingly, in both cases the image shows that the states have sub-nanometer dimension, indicating that they originate from atomic defects (no more specific fingerprints enabling a direct identification of the defects could be obtained on gated CVD-grown MoS$_2$ layers on SiO$_2$, which prevents the direct identification of the defects by STM imaging alone). The states have energy close to the top of the valence band, as illustrated by spectra 2 and 3 on the left side of Figure 5. Specifically, in region 2 the in-gap states lead to a fully developed, broad peak 400 meV above the top of the valence band, and in region 3 in a shoulder at a comparable energy (again 300-400 meV above the top of the valence band). The energy of these features (i.e., their distance from the top of the valence band) coincide with what we observed in measurements of the source-drain current as a function of gate/reference potential (see Figure 1b and Figure 4). From this we conclude the in-gap states created by atomic scale defects are the same states acting as hole traps, which are responsible for the observed anomalous transport properties discussed above.

The combination of all the measurements presented so far (transport experiments on exfoliated and CVD-grown MoS$_2$ monolayers, their comparison, the comparison with the normal ambipolar behavior observed in other semiconducting TDMCs,  photoluminescence and Raman measurements, and the  STM and STS measurements just discussed), in conjunction with theoretical investigations based on {\it ab-initio} calculations reported in the literature\cite{Noh2014,Yuan2014,Huang2015,Mahjouri-Samani2016,Vancso2016}, point to a realistic and fully consistent scenario accounting for the experimental observations. Specifically, we claim that the atomic scale defects that we observe and that are responsible for the anomalous hole transport in exfoliated MoS$_2$ monolayers are individual or clustered sulfur vacancies, which are known to be present in relative high density in MoS$_2$ bulk crystals, and therefore also in exfoliated monolayers. Spectroscopy measurements show that they create in-gap states near the top of the valence band acting as traps and strongly affecting hole transport, which is in agreement with expectations based on  {\it ab-initio} calculations \cite{Yuan2014,Huang2015,Vancso2016}. More specifically, it is known in the context of semiconducting TMDC monolayers that the detailed analysis of the electronic states generated by sulfur and selenium vacancies is complex, as these defects create different types of localized states. However, if we confine our attention exclusively to defect-induced states near the top of the valence band (\textit{i.e.}, those states that can act as effective traps for holes), the results of virtually all existing {\it ab-initio} calculations\cite{Noh2014,Yuan2014,Huang2015,Mahjouri-Samani2016,Vancso2016} indicate that only in the case of MoS$_2$ monolayers these states are inside the band gap. In other semiconducting TMDC monolayers (WS$_2$, MoSe$_2$, and WSe$_2$) the energy of states near the top of the valence band generated by sulfur or selenium vacancies is actually in the band. As a result, in these other monolayers the defect states do not act as hole traps, because they can hybridize with delocalized states at the same energy. That is: sulfur vacancies explain why only in MoS$_2$ monolayers hole transport is anomalous. For MoS$_2$ bilayers or thicker multilayers the absence of any anomaly in hole transport is explained in a similar way: the decrease of the band gap with increasing the number of layers (by at least 0.5 eV for bilayers as compared to monolayers, and more for thicker multilayers) makes the states induced by sulfur vacancies "fall" into the valence band, preventing them from acting as hole traps.

Sulfur vacancies also accounts for other aspects of our observations. In particular, as we have remarked above, the defects affecting hole transport are the same in exfoliated monolayers and CVD-grown ones, albeit in CVD-grown monolayers their effect is less intense. Sulfur vacancies are compatible with this conclusion, as they are expected to be present in both systems. This is not the case for most other types of atomic defects: indeed, mineral MoS$_2$ crystals are known to contain a number of different impurities \cite{Addou2015a}, but these same impurities are neither necessarily present in the Mo and S source material used for CVD growth, nor would they be effectively transferred to the monolayer during the growth process if present. That is why finding that the defects responsible for hole trapping are the same in exfoliated and CVD-grown monolayers strongly constraints the number of possibilities, and sulfur vacancies is one of the few. Finally, sulfur vacancies also naturally explain why the effect of disorder on hole transport is much less pronounced in devices fabricated on CVD-grown material, simply because under the conditions at which CVD growth is done a large excess of sulfur is present, which leads to largely reduced density of sulfur vacancies as compared to that present in exfoliated flakes.

\section*{Conclusion}
Simple theoretical considerations suggest that monolayers of group VI semiconducting transition metal dichalcogenides all possess very similar electronic properties and functionality. Inasmuch as this may be the case for ideal defect-free systems, actual materials contain defects whose influence on the physical properties  can (and do) strongly depend on the compound considered. Here we have shown that this is clearly the case for hole transport, which in monolayer MoS$_2$ -- but not in monolayers of other common TDMCs -- is strongly affected by defects naturally present. Our results very strongly support the conclusion that these naturally present defects are sulfur vacancies, and explain why these defects have a different effect in MoS$_2$ monolayers as compared to monolayers of other similar compounds, \textit{i.e.} the fact that for MoS$_2$ monolayers sulfur vacancies cause localized defect states at energies inside the band gap. They also illustrate a possible solution to minimize the influence of these defects on the performance of devices realized using monolayer MoS$_2$, namely the use of material grown in a large excess of sulfur. \\

There is a broad consensus that 2D semiconducting materials have an important potential for technology in the long term. It is clear that exploiting this potential cannot only rely on properties of the idealized systems, but has to take into account the effect of defects unavoidably present, understand it in detail, and find strategies to minimize it. Our results illustrate how this can be done in a concrete specific case.

\section*{Methods}

MoS$_{2}$ flakes are exfoliated from commercially available crystals (SPI supplies and 2D Semiconductors) by conventional scotch tape exfoliation technique. CVD monolayers of MoS$_{2}$ are synthesized through chemical reaction of a molybdenum containing precursor (MoO$_{3}$) and sulfur, according to a protocol reported in ref.45. In short, crucibles with the solid precursors are loaded into a tube furnace so that the MoO$_{3}$ powder is positioned in the middle of the furnace and the sulfur approximately 20 cm upstream. The substrate (either sapphire or SiO$_{2}$) is mounted 'face-down' above the MoO$_{3}$-containing crucible. The synthesis is done under constant Ar flow of 75 sccm. Throughout the growth the temperatures of MoO$_{3}$ and sulfur are maintained at 700 and 250$^{\circ}$C respectively.

Electronic transport measurements are done using FETs with ionic liquid top gate. Electric contacts are fabricated with electron-beam lithography, metal evaporation, and lift-off. Devices are further annealed at 200$^{\circ}$C in an inert atmosphere of Ar for 2 hours in order to decrease the contact resistance. A drop of ionic liquid (P14-FAP 1-butyl-1-methylpyrrolidiniumtris(pentafluoroethyl)trifluorophosphate) is put on top of the devices right before loading them into a vacuum chamber (p$\thickapprox$10$^{-6}$ mbar). Prior to measurements devices are kept in the chamber overnight to pump out humidity and oxygen.

The STM experiments were done in ultra-high vacuum (base pressure 5$\cdot$10$^{-12}$ mbar) at liquid nitrogen temperature using tips electrochemically etched from an annealed tungsten wire. Topographic images were recorded in constant current mode. Tunneling spectroscopic I(V) curves were acquired with open feedback loop while $\mathbf{d}I/\mathbf{d}V$ curves were obtained by numerical derivation of the measured I(V) curves. Positive and negative bias voltages correspond to empty and filled states of the sample, respectively.

Photoluminescence and Raman measurements were performed both in back-scattering geometry (\textit{i.e.} collecting the emitted light with the same microscope used to couple the laser beam onto the device). The light collected from the sample was sent to a Czerny-Turner monochromator and detected with a Si CCD-array (Andor). For both type of measurements the excitation wavelength of the laser was set to 514.5 nm and the power was kept below 20 $\mu W$.

\section*{Acknowledgments}

We are grateful to A. Ferreira, R. Villarreal, and G. Manfrini for fruitful discussions and technical help. AFM and CR gratefully acknowledge financial support from the Swiss National Science Foundation (Division 2 and Sinergia program). AFM also gratefully acknowledges financial support from the EU Graphene Flagship project. NU acknowledges funding from an Ambizione grant of the Swiss National Science Foundation.

\section*{Associated content}
Supporting Information Available: FET transfer curves with Pt and Au contacts, photoluminescence map, FET transfer curve at reduced temperature. This material is available free of charge via the Internet at http://pubs.acs.org.

\section*{Competing financial interests}
The authors declare no competing financial interests.

\providecommand{\latin}[1]{#1}
\makeatletter
\providecommand{\doi}
{\begingroup\let\do\@makeother\dospecials
	\catcode`\{=1 \catcode`\}=2 \doi@aux}
\providecommand{\doi@aux}[1]{\endgroup\texttt{#1}}
\makeatother
\providecommand*\mcitethebibliography{\thebibliography}
\csname @ifundefined\endcsname{endmcitethebibliography}
{\let\endmcitethebibliography\endthebibliography}{}

\end{document}